# AN INVESTIGATION OF OPTIMISED FREQUENCY DISTRIBUTIONS FOR DAMPING WAKEFIELDS IN X-BAND LINACS FOR THE NLC

R.M. Jones[1], SLAC; N.M. Kroll[2], UCSD & SLAC; R.H. Miller[1],
T.O. Raubenheimer[1] and G.V. Stupakov[1]; SLAC

*Abstract*

In the NLC (Next Linear Collider) small misalignments in each of the individual accelerator structures (or the accelerator cells) will give rise to wakefields which kick the beam from its electrical axis. This wakefield can cause BBU (Beam Break Up) or, at the very least, it will dilute the emittance of the beam. Several Gaussian detuned structures have been designed and tested [1] at SLAC and in this paper we explore new distributions with possibly better damping properties. The progress of the beam through approximately 5,000 structures is monitored in phase space and results on this are presented.

## 1. INTRODUCTION

In all of our previous accelerating structures [2], the cell dimensions have been designed such that they follow an Erf function profile and the uncoupled cells have a Gaussian Kdn/df, kick-factor weighed density function,

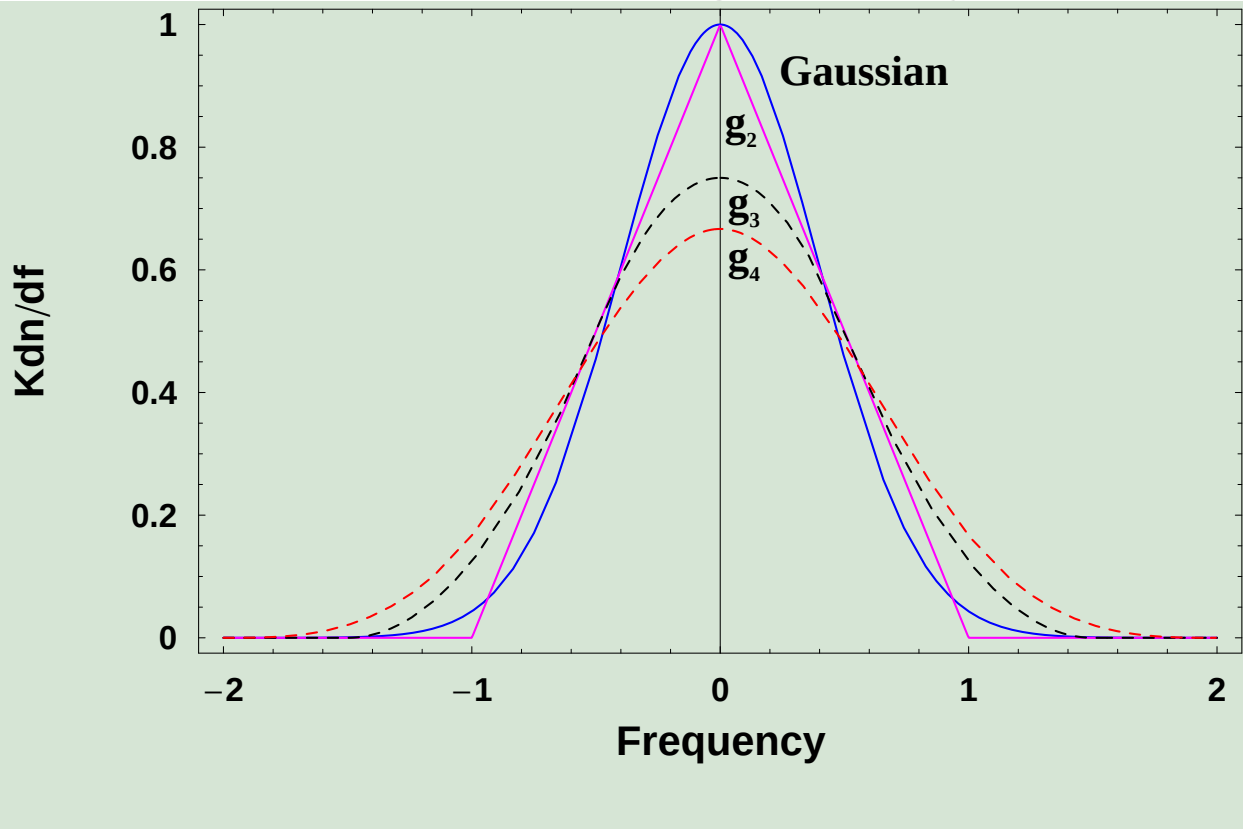

Figure 1: Optimisation with the idealised frequency distributions illustrated

[3] profile. The normalised Gaussian distribution is shown in fig. 1 together with the convolution of a number of rectangular distributions. An ideal Gaussian falls off as $\exp(-f^2)$. However, in practise, a Gaussian distribution of the uncoupled modes leads to a wakefield which does not continue to fall rapidly because of three effects. Firstly, due to the truncation of the Gaussian distribution (this gives rise to a 1/s decay), secondly, due to sampling the Gaussian function (with a finite number of cells and a specified frequency bandwidth) and thirdly, cell-to-cell coupling perturbs the original Gaussian distribution function. All three effects reduce the rate of decay.

In this paper we address the issue of the truncation of the Kdn/df distribution. The envelope wakefield for a truncated Gaussian function (shown in fig. 2) only follows a Gaussian decay for the initial part of the decay (the first few bunches) and thereafter a considerable 'ripple' occurs. Additional moderate damping (Q~1000) is generated with four manifolds that lie along the outer wall of the cavities but this only takes effect after 10's of nanoseconds. Thus, these 'ripples' can have serious consequences on the wakefield.

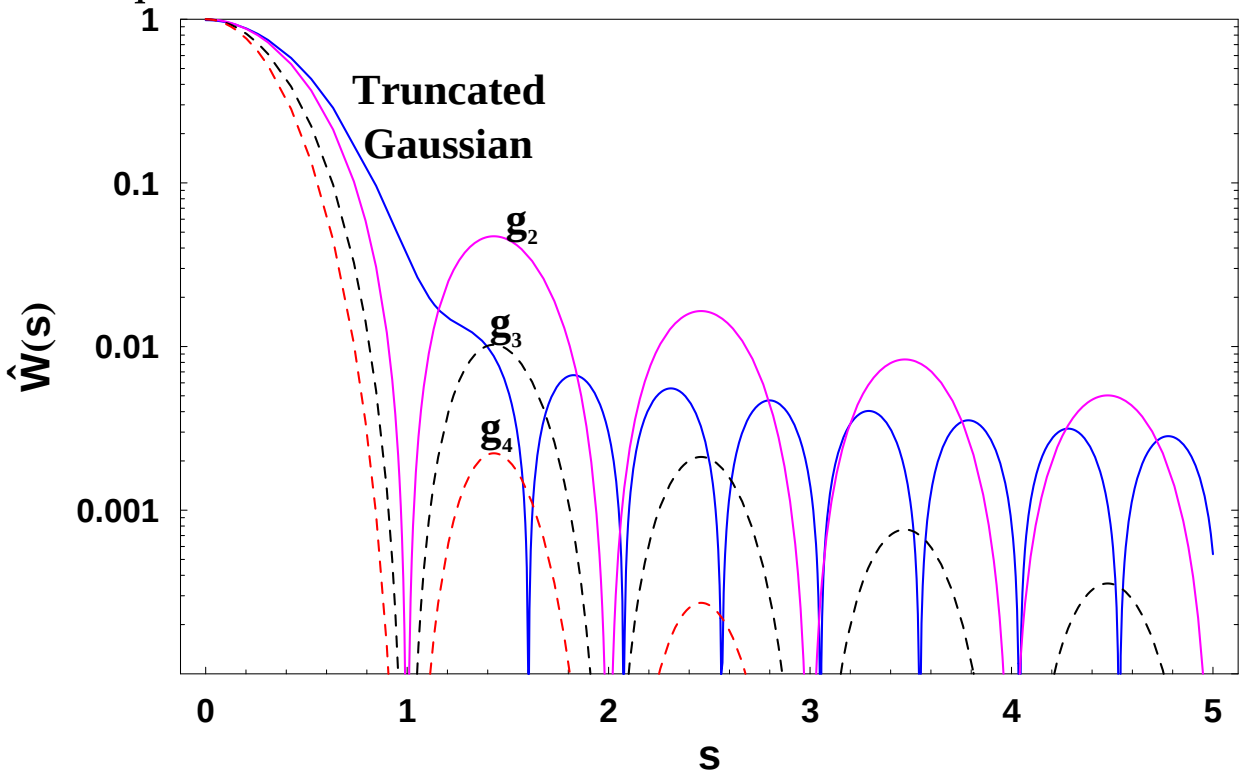

Figure 2: Wake function for idealised distributions

In order to reduce the large 'ripple' we have considered various distributions to replace the Gaussian prescription. Here, we will concentrate on a number of convolutions of rectangular distributions. A rectangular distribution has a sinc function as its Fourier transform. The sinc has zeros at s=n and lobes which fall off as 1/s. Each additional convolution leads to a $1/s^k$ fall-off in the wakefield. It is interesting to note that in the limit of an infinite number of self-convolutions the resulting distribution approaches a Gaussian distribution and this arises as a consequence of the central limit theorem [4]. Here, we consider k=2 (a triangular distribution, $g_2$) and k=3 (the convolution of a triangular function with a top hat function, $g_3$) and k=4 the self-convolution of the triangular function and these are shown in fig 1. The Fourier transform of the k=4 case is given by $\text{sinc}^4$ function and this is compared with the truncated Gaussian in fig. 2. together with the k=2 and k=3 cases. The function described by the k=4 case is identically zero at frequency units $\pm 2$ and thus enforced truncation is not necessary. The peak values in the ripples of the wakefield of the truncated Gaussian lie below the sinc$^2$ but not below the sinc$^4$ function. For this reason we choose a $g_4$ (sinc$^4$ in wake space) design for a new RDDS based upon a mapping function [5] re-parameterisation of RDDS1.

[1] Supported under U.S. DOE contract DE-AC03-76SF00515
[2] Supported under U.S. DOE grant DE-FG03-93ER407

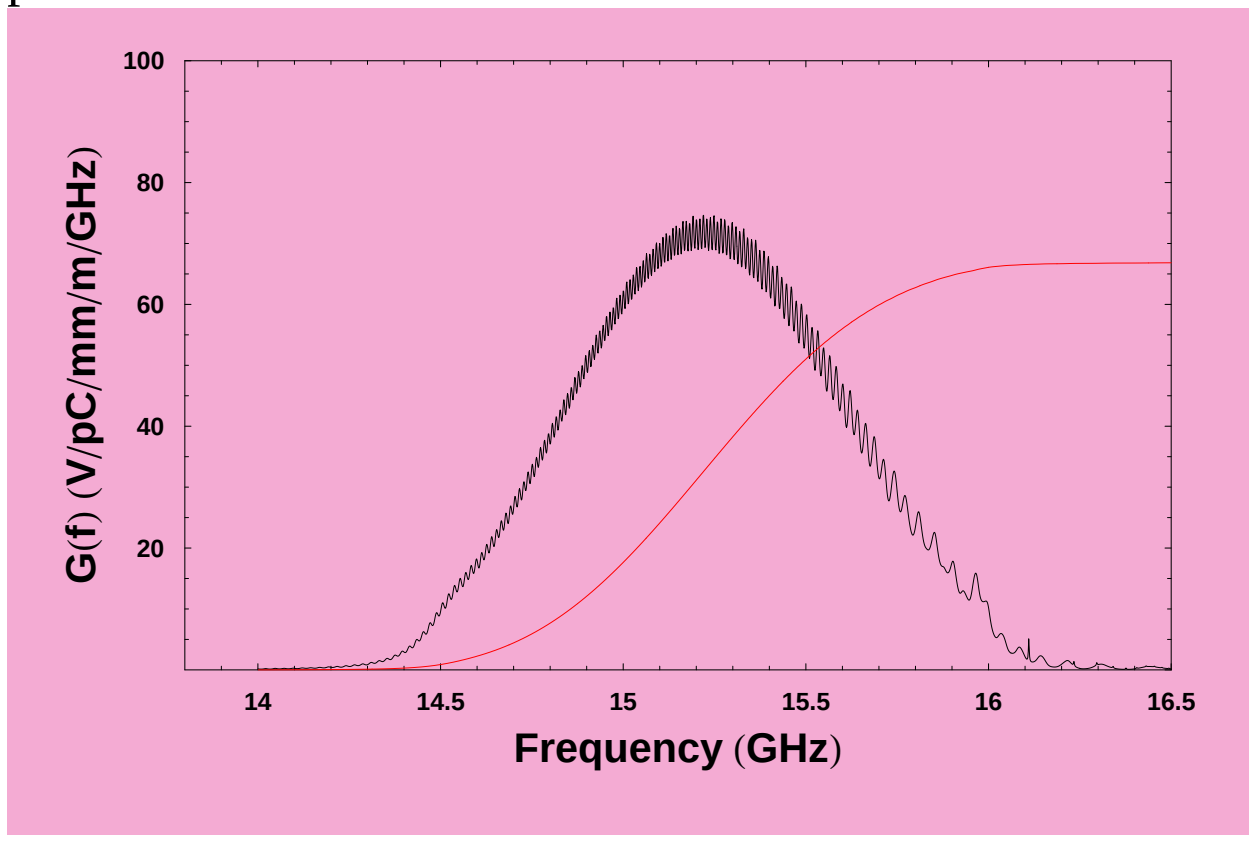

Figure 3: G(f), Spectral function, for a sinc$^4$ variation

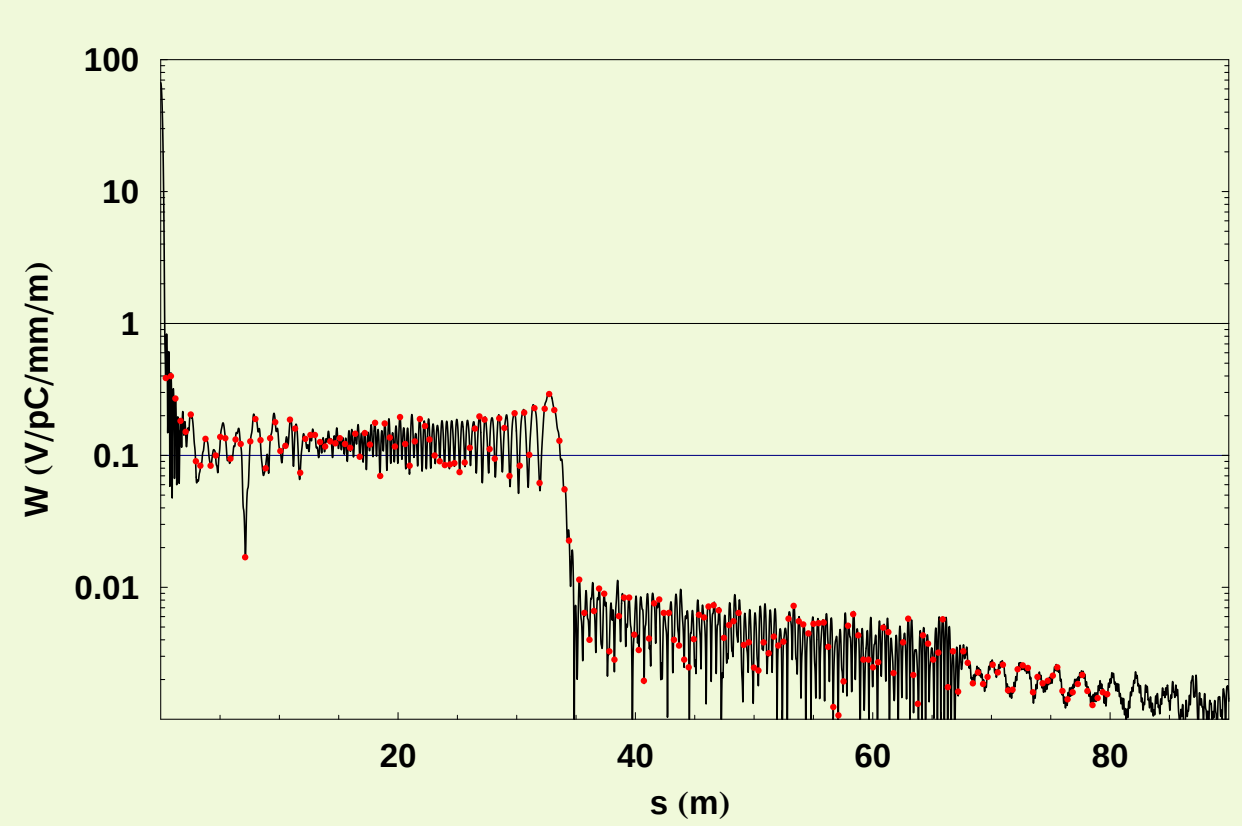

Figure 4: Wakefield for Sinc$^4$ distribution

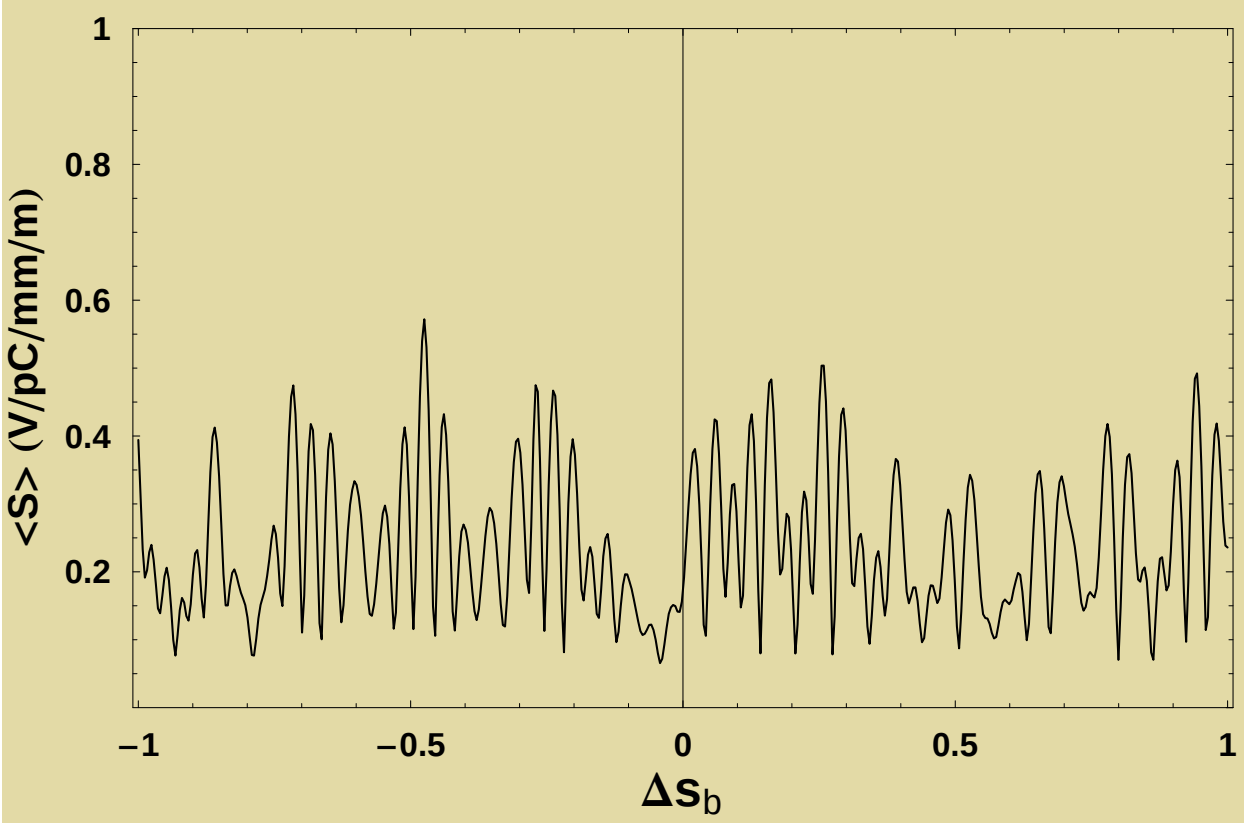

Figure 5: Sum wake function for optimised distribution

## 2. WAKE FOR A SINC$^4$ DISTRIBUTION

We compute the wake envelope function using the spectral function method [6] which has proven accurate in predicting the wakefield in a measured structure [1]. The spectral function for sinc$^4$ is shown in fig. 3. The main difference between it and the spectral function of RDDS1 (fig 6) lies in the upper frequency end of the distribution. In the case of RDDS1, the kick factors increase almost linearly with synchronous frequency and towards the high frequency end of the first dipole band the mode density (dn/df) has to increase in order that Kdn/df be a symmetric function that falls with a Gaussian profile. However, as dn/df increases then, of course, the modal separation increases to approximately 50 MHz or more as compared to 7MHz in the center of the band With the large separation, the modes are not particularly well damped by the manifold in the high frequency region. However, the sinc$^4$ possesses the useful property that the modes are much more well damped in this region (15.8 GHz and beyond).

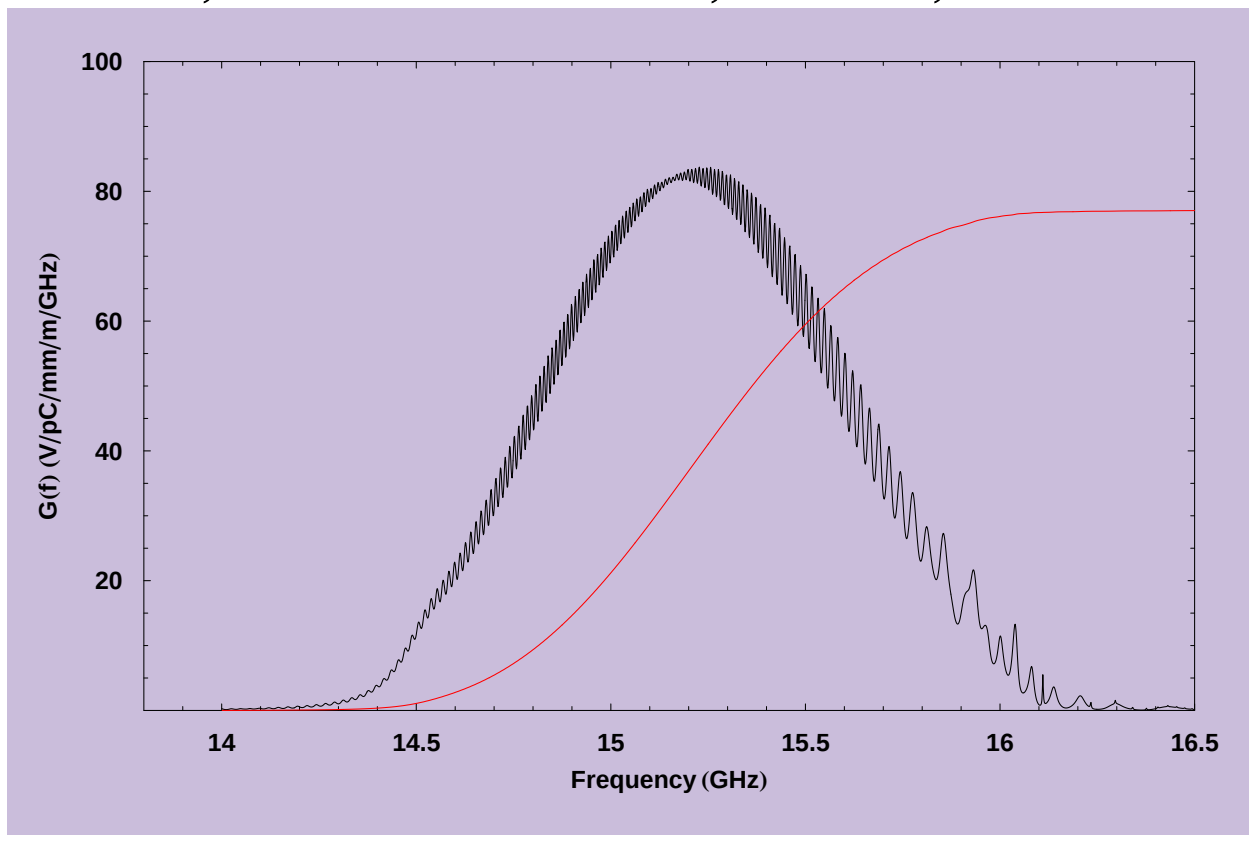

Figure 6: Spectral function RDDS1

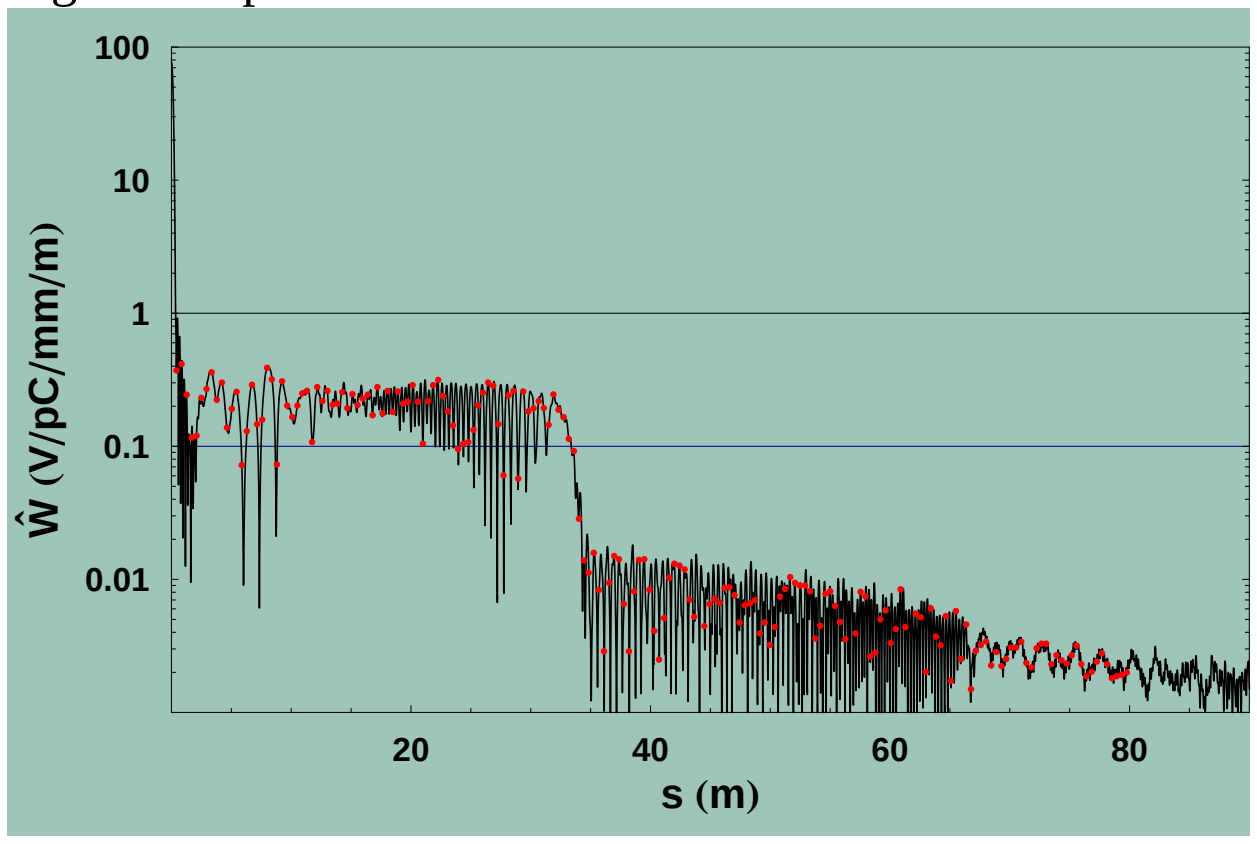

Figure 7: Envelope of wakefunction for RDDS1

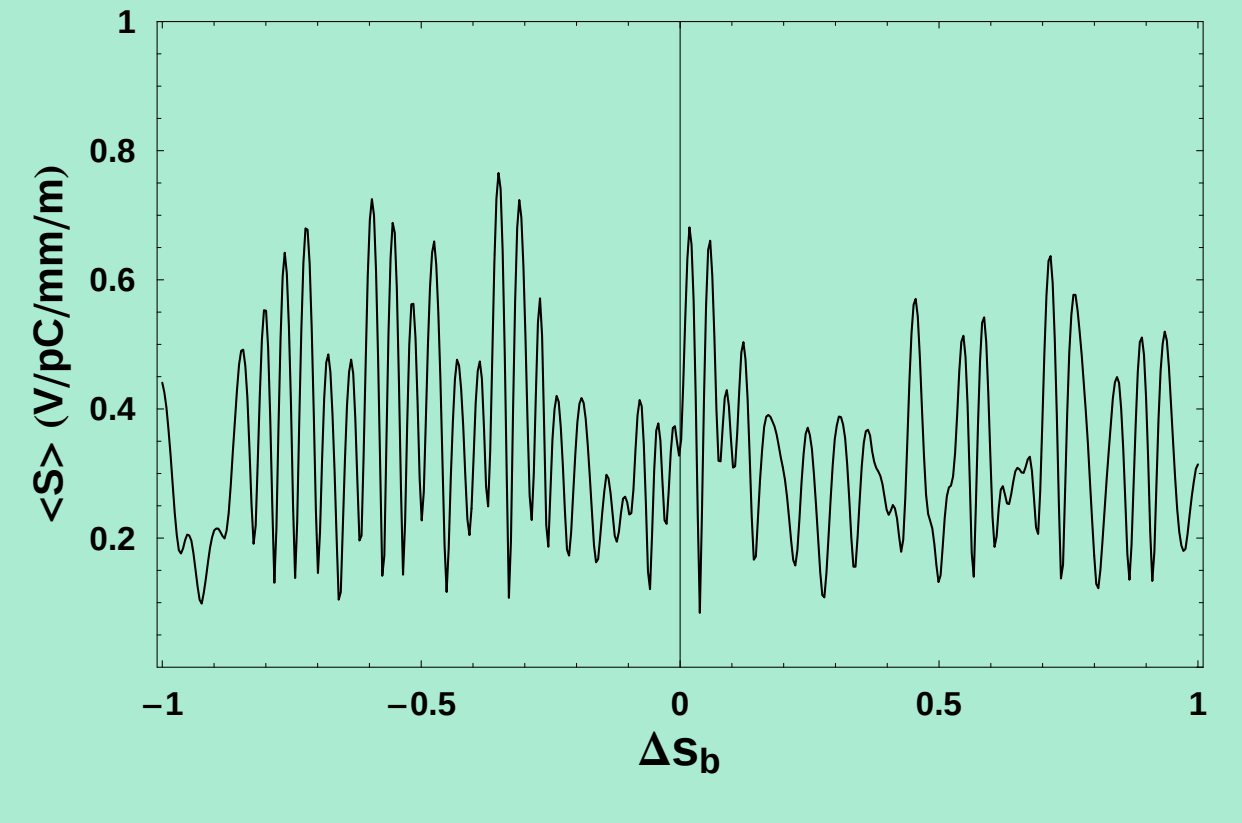

Figure 8: Sum wake function for RDDS1

The wakefield corresponding to the spectral function of fig 3, is shown in fig 4 and the main improvement over the wakefield of our present structure, RDDS1 (shown in fig. 7) lies in the region 0 to 10 m in which the wakefield is improved by a factor of approximately 2 or more. Also shown in figs 5 and 8 is $S_\sigma$, the standard deviation of the sum wakefield from the mean value, for the sinc$^4$ distribution and RDDS1 respectively, as a function of $\Delta S_b$, the relative variation in the bunch spacing (expressed as a percentage). The sum wakefield is useful in that it provides an indicator as to whether or not BBU will occur. The abscissa in these curves is $\Delta S_b$ and this provides a convenient means of shifting all the cell frequencies by a fixed amount and it corresponds to a systematic error in the synchronous frequencies. From previous simulations, peaks in the standard deviation of the sum wakefield close to unity have proved to be a symptom of BBU. However, BBU is indeed a complex phenomena and, in order to be sure that BBU will actually take place many particle tracking simulations with the code LIAR [7] need to be undertaken. In the next section the results on particle tracking simulations at peak values in $S_\sigma$ are presented.

## 3. BEAM DYNAMICS: TRACKING THROUGH COMPLETE LINAC

In all of the tracking simulations we performed the bunch train is offset by 1μm and its progress down the linac is monitored. Additional details regarding the simulation parameters are given in [8]. At the nominal bunch spacing (84 cm) $S_\sigma$ is approximately 0.15 V/pC/mm/m and 0.3 V/pC/mm/m for the new distribution and for RDDS1, respectively. Tracking through the complete linac for both distributions indicates that that no significant emittance dilution arises. In both cases, there are peaks in $S_\sigma$ are very close (less than .05%) to the nominal bunch spacing, however simulations show that these also give rise to no more than 1 or 2 percent dilution of the beam emittance. The largest peak in $S_\sigma$ for RDDS1 and the new distribution are located at -0.35 and –0.48% away from the nominal bunch spacing, respectively. The emittance growth after tracking through the linac at these modified bunch spacings is shown fig. 9. For the sinc$^4$ distribution there is no emittance dilution arising from long range wakes. However, approximately 6% emittance growth occurs for RDDS1. The phase space, shown in fig 10, indicates that for the sinc$^4$ distribution the particles are well contained but for RDDS1 the bunch train is starting to break up. Nonetheless, emittance growth is unlikely to be a problem for RDDS1 because: firstly the the systematic shift is unlikely to be so large (-0.48% in the bunch spacing corresponds to a shift in the dipole mode frequency of 72 MHz) and secondly, the shift is not expected to be identical from structure-to-structure which has been shown [6] to significantly reduce the BBU.

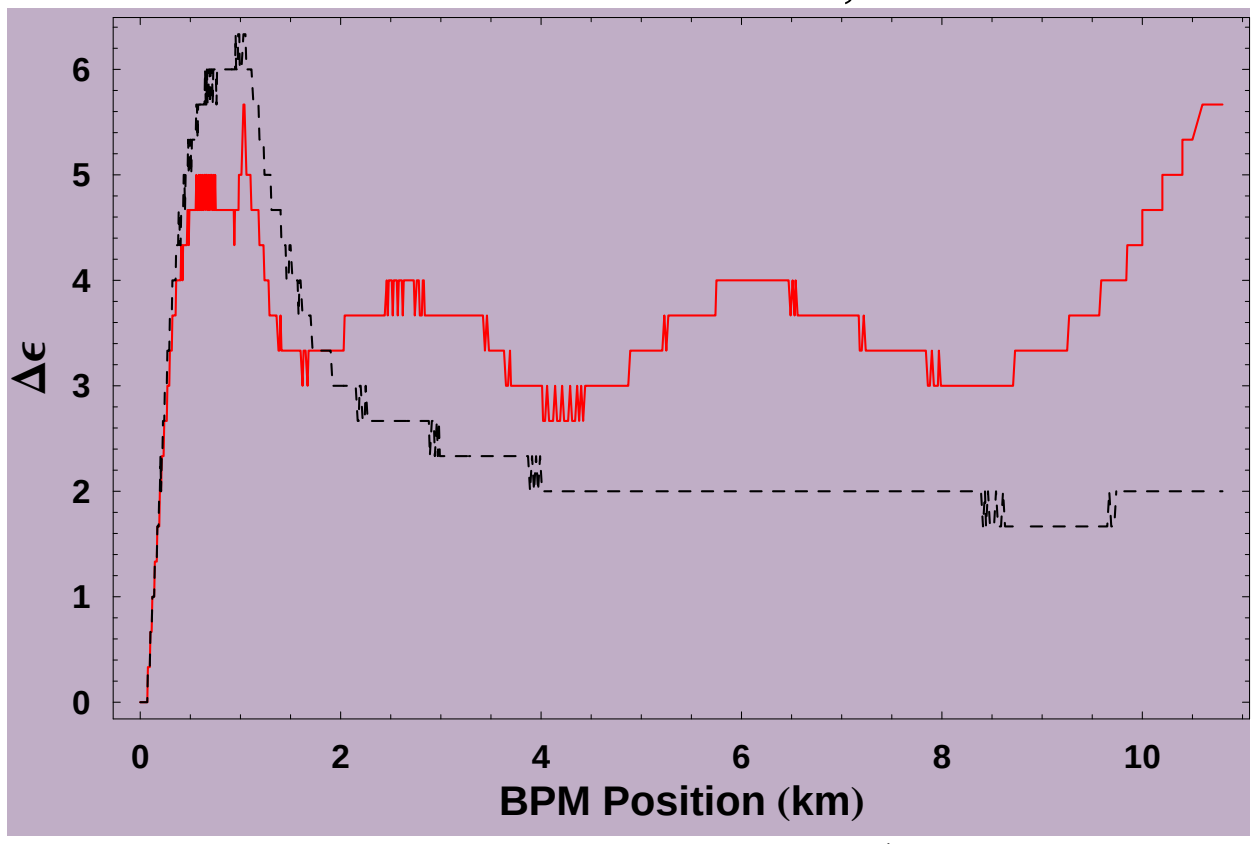

Figure 9: Emittance growth for the sinc$^4$ distribution and RDDS1 at a bunch spacing which maximises $S_\sigma$

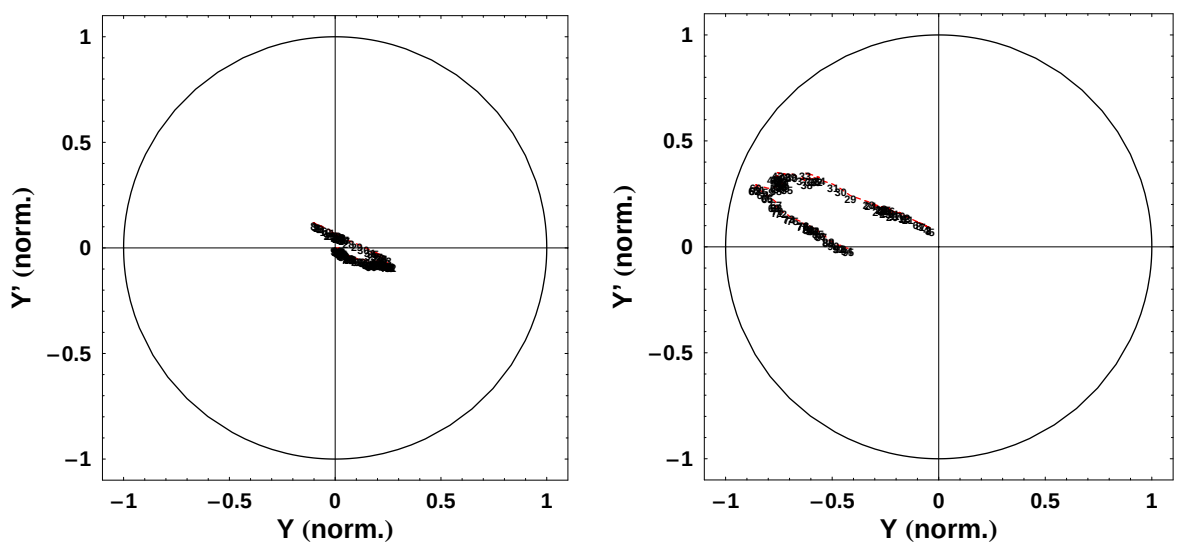

(a) (b)

Figure 10: Phase space for sinc$^4$ distribution (a) and RDDS1 (b) at a bunch spacing which maximises $S_\sigma$

## 4. CONCLUSIONS

A sinc$^4$ distribution for the uncoupled distribution leads to improved damping of the transverse wakefield. The mean value of $S_\sigma$ is approximately 2 times smaller than that of our present structure, RDDS1 and, we have found that no significant emittance growth occurs over a broad range of systematic shifts in the synchronous frequencies of the cells. However, additional optimisation of the frequency distribution and in the coupling of the wakefield to the manifold, should lead to even better damping of the wakefield. In the near future, we plan to embark on a program of iterative optimisation of the wakefield in which the uncoupled distribution is modified to obtain the desired coupled distribution. This should allow us to take full advantage of the uniformly spaced zeros of the sinc distribution. We would then choose the lobe spacing to equal the bunch spacing, which should yield a very small wakefield.